\documentclass[10pt,twocolumn,superscriptaddress,preprintnumbers,showpacs]{revtex4}
\usepackage{epsf}
\usepackage{amsmath}
\usepackage{amssymb}
\usepackage{epsfig}
\usepackage{latexsym}
\usepackage{amsfonts}
\usepackage{varioref}
\usepackage{ifthen}
\usepackage{graphicx}
\usepackage{amssymb,amsmath,amsthm}%
\providecommand{\U}[1]{\protect\rule{.1in}{.1in}}
\begin{document}
\parindent 0mm 
\parindent 0mm 
\setlength{\parskip}{\baselineskip}
\pagenumbering{arabic} 
\setcounter{page}{1}
\mbox{ }
\preprint{UCT-TP-292/12}
\title{Rho-meson resonance broadening in QCD at finite temperature}
\vspace{.1cm}
\author{Alejandro Ayala}
\affiliation{Instituto de Ciencias Nucleares, Universidad Nacional
Aut\'onoma de M\'exico, Apartado Postal 70-543, M\'exico Distrito Federal
04510, Mexico}
\author{C. A. Dominguez}
\affiliation{Centre for Theoretical \& Mathematical Physics, University of Cape Town,
Rondebosch 7700, South Africa}
\author{M. Loewe}
\affiliation{Centre for Theoretical \& Mathematical Physics, University of Cape Town,
Rondebosch 7700, South Africa}
\affiliation{Facultad de F\'{i}sica, Pontificia Universidad Cat\'{o}lica de Chile, Casilla 306, Santiago 22, Chile}
\author{Y. Zhang}
\affiliation{Centre for Theoretical \& Mathematical Physics, University of Cape Town,
Rondebosch 7700, South Africa}
\date{\today}
\begin{abstract}
Thermal Finite Energy QCD sum rules for the vector current correlator are used to study quark-gluon deconfinement. Assuming $\rho$-meson saturation of the correlator in the hadronic sector, and the Operator Product Expansion in QCD, we obtain the temperature behavior of the resonance parameters (coupling, mass, and width), and of the leading vacuum condensates, as well as the perturbative QCD threshold in the complex squared energy plane. The results are consistent with quark-gluon deconfinement at a critical temperature $T_c \simeq 197\; {\mbox{MeV}}$. The temperature dependence of the $\rho$-meson width is of importance for current experiments on dimuon production in nuclear collisions.
\end{abstract}
\maketitle
\section{INTRODUCTION}
The temperature behavior of the light-quark vector current correlator in the framework of thermal QCD sum rules (QCDSR) was first discussed in \cite{BS}, and later reanalyzed by others \cite{Rho} mostly using Laplace transform sum rules. Inconsistent results from the use of these sum rules were first pointed out in \cite{CAD1}-\cite{CAD2}. A better understanding of the QCDSR  method \cite{REVIEW}, both at zero and at finite temperature, and in particular the use of Cauchy's theorem in the complex energy plane to formulate quark-hadron duality, has led to a preference of  Finite Energy QCDSR (FESR) over the Laplace transform counterparts. In fact, modern determinations of the QCD strong coupling \cite{Pich} and quark masses \cite{CAD3}, as well as thermal properties of hadrons \cite{CAD4a}-\cite{CAD4} are now mostly done in the framework of FESR. Regarding the latter, the emerging picture is as follows. A key parameter signaling quark-gluon deconfinement is the squared energy threshold, $s_0(T)$, above which the hadronic spectral function is well approximated by perturbative QCD (PQCD), as first proposed in \cite{BS}. Explicit determinations of $s_0(T)$ in several light- and heavy-light-quark systems \cite{BS}-\cite{CAD2}, \cite{CAD4a}-\cite{CAD4} find this parameter to decrease with increasing temperature, vanishing at a critical value $T=T_c$ interpreted as the deconfinement temperature. Two other hadronic parameters, the coupling and the width, also behave as expected from a deconfinement scenario, i.e. the coupling decreases and the width increases with increasing temperature. A monotonically increasing hadronic width,
interpreted as resonance absorption in a hot plasma, was proposed long ago as a clear signal of deconfinement \cite{BROAD}, and used to predict resonance broadening in dimuon production in high energy nuclear collisions \cite{DIMUON}. This effect was later confirmed by several experiments \cite{BROADEXP}.
An exception to the above behaviour has been found for the heavy-heavy-quark states $J/\psi$, $\eta_c$, and $\chi_c$, which appear to survive above $T_c$ \cite{CAD5} in agreement with lattice QCD results \cite{LQCD}. It should be recalled that the conceptual interpretation of a hadronic width at finite temperature is different from that at $T=0$. While in the latter case the width is entirely related to decay into allowed hadronic channels, at $T \neq 0$ particles hadronically stable at $T=0$ develop a width due to the emergence of scattering channels which  modify the interaction rate. This has been demonstrated explicitly in hadronic models, e.g. the linear sigma model \cite{sigma}, chiral perturbation theory \cite {CHPTT}, and others, as well as in QCD sum rules \cite{CAD4a} , where e.g. the pion and the nucleon develop a width which increases monotonically with temperature. In other words, an increasing width signals  an increase in the interaction probability, rather than only a decrease in a decay rate.
\\ 
Regarding the hadron mass, its temperature behavior is hardly of any interest in the context of the deconfinement transition. In fact, the mass is nothing but the real part of the pole of the hadron propagator in the complex squared energy plane. Whether the pole moves up or down with increasing $T$ does not provide, in itself, any information regarding deconfinement. It is the imaginary part of the hadron propagator, i.e. the hadronic width, which signals deconfinement if it grows with increasing $T$. One could picture the extreme situation of the mass decreasing and vanishing at some $T^*$. If this behavior is not accompanied by an increasing width then the hadron would still be present in the spectral function at $T=T^*$. Explicit determinations of the temperature dependence of hadron masses indicate a mild change with temperature, increasing or decreasing slightly depending on the channel.\\
In this paper we consider the first three thermal FESR for the  vector current correlator  to find the behavior of $s_0(T)$ and the $\rho$-meson parameters, as well as the dimension $d=4$ gluon condensate, and the $d=6$ four-quark condensate entering the Operator product Expansion (OPE) in QCD.  This follows a recent analysis \cite{CAD6} of the axial-vector channel with a much improved hadronic spectral function, involving not only the pion pole as in \cite{BROAD}, \cite{Axial} but also the $a_1 (1260)$ resonance. Since the pion pole in this channel is related to the quark condensate, a connection can be established between chiral-symmetry restoration and deconfinement. Given the absence of a pole in the vector channel, it is not a-priori obvious that results in this case should be similar to those in the axial-vector channel, where the thermal quark condensate mostly drives the behaviour of $s_0(T)$. Another difference between axial-vector and vector channels is the presence in the latter of a space-like cut in the complex energy plane, interpreted as due to the scattering of the vector current off pions in the hot plasma \cite{BS}. This term is of higher order (two-loop) in the axial-vector channel, and thus negligible \cite{BROAD}, \cite{CAD6}-\cite{Axial}. On the other hand, PQCD is chiral symmetric (in the chiral limit), and so is the leading dimension $d=4$ term in the OPE. Additional motivation for reexamining the thermal vector current correlator using FESR, and modern information, is to determine the $T$-dependence of the $\rho$-meson width entirely from QCD, as opposed to earlier work based on current algebra \cite{DIMUON}. This is important in connection with current experiments on dimuon production in high energy nuclear collisions \cite{BROADEXP}. 
\begin{figure}[hb]
\centering
\def\svgwidth{0.8\columnwidth}
\includegraphics[height=3.0in, width=3.5in]{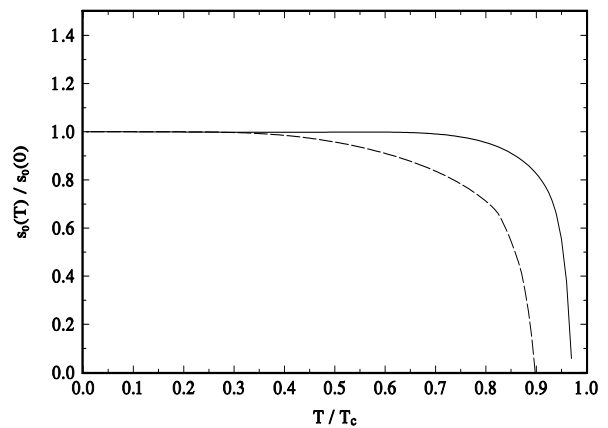}
\caption{{\protect\small{The normalized thermal behavior of the PQCD threshold in the vector channel (solid curve), for $T_c = 197 \;{\mbox{MeV}}$, and in the axial-vector channel (dotted curve) from \cite{CAD6} for the same $T_c$.}}}
\end{figure}
\begin{figure}
\centering
\def\svgwidth{0.8\columnwidth}
\includegraphics[height=3.0in, width=3.5in]{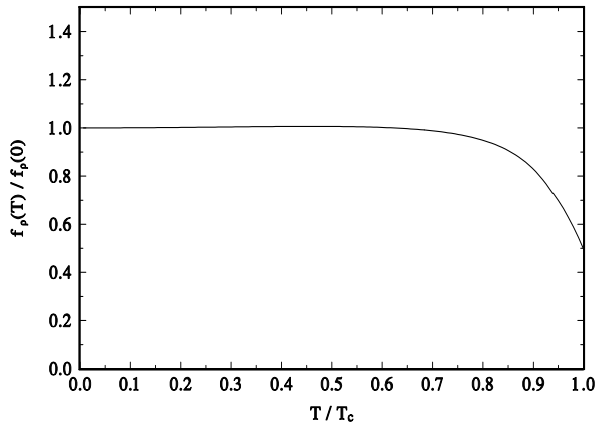}
\caption{{\protect\small{The normalized thermal behavior of the $\rho$-meson leptonic decay constant, for $T_c = 197 \;{\mbox{MeV}}$.}}}
\end{figure}
\section{QCD SUM RULES}
The light-quark vector current correlator at $T=0$ is defined as
\begin{eqnarray}
\Pi_{\mu\nu} (q^{2})   &=& i \, \int\; d^{4} \, x \, e^{i q x} \,
<0|T( V_{\mu}(x)   V_{\nu}^{\dagger}(0))|0> \nonumber \\ [.3cm]
&=& (-g_{\mu\nu}\, q^2 + q_\mu q_\nu) \, \Pi(q^2)  \; ,
\end{eqnarray}
where $V_\mu(x) = \frac{1}{2}[: \bar{u}(x) \gamma_\mu \, u(x) - \bar{d}(x) \gamma_\mu \, d(x):]$ is the (electric charge neutral) conserved vector current in the chiral limit, and $q_\mu = (\omega, \vec{q})$ is the four-momentum carried by the current. The function $\Pi(q^2)$ in PQCD is normalized as
\begin{equation}
	{\mbox{Im}} \,\Pi(q^2) = \frac{1}{8\pi}\left[ 1 + {\cal{O}} \left( \alpha_s(q^2)\right)\right],
\end{equation}
where radiative corrections are currently known up to five-loop order, i.e. ${\cal{O}}(\alpha_s^4)$.
The QCD FESR rest on two pillars \cite{REVIEW}, (i) the operator product expansion (OPE) of current correlators at short distances beyond perturbation theory, and (ii) Cauchy's theorem in the complex squared energy plane, which relates the (hadronic) discontinuity across the cut on the real semi-axis with the integral around a contour of radius $|s_0|$ where the OPE is expected to be valid. The latter is usually referred to as quark-hadron duality. This leads to the FESR
\begin{eqnarray}
&(-)^{(N-1)}& C_{2N} \langle {\mathcal{\hat{O}}}_{2N}\rangle = 8 \pi^2 \int_0^{s_0} ds\, s^{N-1} \,\frac{1}{\pi} {\mbox{Im}} \Pi(s)|_{\mbox{\scriptsize
{HAD}}}
\nonumber \\ [.3cm]
&-& \frac{s_0^N}{N} \left[1+{\mathcal{O}}(\alpha_s)\right] \;\; (N=1,2,\cdots) \;,\label{FESR}
\end{eqnarray}
where the leading order vacuum condensates in the chiral limit are the dimension $d=4$ gluon condensate
\begin{equation}
C_4 \langle \hat{\mathcal{O}}_{4}  \rangle = 
\frac{\pi}{3} \langle \alpha_s G^2\rangle  , \label{C4}
\end{equation}
and the dimension $d=6$ four-quark condensate
\begin{equation}
C_6 \langle \hat{\mathcal{O}}_{6}  \rangle = - 8 \pi^3 \alpha_s \left[\langle (\bar{q} \gamma_\mu \gamma_5 \lambda^a q)^2 \rangle + \frac{2}{9} \langle(\bar{q} \gamma_\mu \lambda^a q)^2\rangle \right]
. \label{C6}
\end{equation}
The radiative corrections in Eq.(3) are known up to five-loop order, i.e. ${\cal{O}}(\alpha_s^4)$, and they will be used at $T=0$ to normalize the FESR.\\
Under the extreme approximation of vacuum saturation, the four-quark condensate can be related to the square of the quark condensate. However, there is no convincing theoretical support for such an approximation. In fact, determinations of the vacuum condensates from data on hadronic decays of the $\tau$-lepton \cite{taucond}, as well as $e^- e^+$ annihilation into hadrons \cite{eecond}, indicate  strong deviations from vacuum saturation. Theoretical arguments from chiral perturbation theory also do not support this approximation at next to next to leading order and at $T=0$  \cite{VS1}. An extension of this analysis to finite temperature indicates a breakdown of vacuum saturation except in the chiral limit \cite{VS2}. 
Since there are no gauge invariant operators of dimension $d=2$ in QCD, it is standard practice to assume they are not present in the OPE. This assumption is supported by results from data analyses \cite{taucond}-\cite{eecond}.\\
In the hadronic sector, assuming $\rho$-meson saturation of the spectral function, and a Breit-Wigner resonance form gives
\begin{equation}
	\frac{1}{\pi} {\mbox{Im}} \Pi|_{HAD}(s) = \frac{1}{\pi} \frac{1}{f_\rho^2} \frac{M_\rho^3 \Gamma_\rho}{\left(s - M_\rho^2\right)^2 + M_\rho^2 \Gamma_\rho^2}\;, \label{BW}
\end{equation}
where \cite{PDG}  $f_\rho = 5$ is the leptonic decay constant, $M_\rho = 0.776$ GeV and $\Gamma_\rho = 0.145$ GeV are the $\rho$ mass and width, respectively. This finite-width parameterization has been normalized such that the area under it equals the area under a zero width expression, i.e. ${\mbox{Im}} \,\Pi|_{HAD}(s) = f_\rho^2 \,M_\rho^2 \,\delta(s - M_\rho^2)$. \\
A test of the FESR, Eq.\eqref{FESR}, with $N=1,2,3$ can be performed by determining $s_0$ together with the gluon condensate and the four-quark condensate, and comparing them with results from data analyses. 
Using the full PQCD information on $\Pi(s)$ to five-loop order, with $\alpha_s(M_\tau^2) = 0.338 \pm 0.012$ \cite{Pich}, and Eq.\eqref{BW} one finds $s_0 = 1.44 \,{\mbox{GeV}}^2$, $C_4\langle{\hat{O}_4}\rangle=0.12$ GeV$^4$ and $C_6\langle{\hat{O}_6}\rangle = - 0.39$ GeV$^6$. These results are in reasonable agreement within errors  with \cite{taucond} and \cite{eecond}. This is not surprising, as the condensate determinations based on experimental data \cite{taucond}-\cite{eecond} require similar values of $s_0$. The value $\sqrt{s_0} = 1.2 \,{\mbox{GeV}}$ validates the assumption of $\rho$-dominance, as the first radial excitation of the $\rho$ is the $\rho(1450)$ with a mass $M \simeq 1.5 \,{\mbox{GeV}}$.\\
The finite-width parametrization, Eq. (6), is clearly not unique. In order to test its impact on the results at $T=0$ we used instead of the standard $\rho$-propagator $1/\left[s - M^2 + i M \Gamma\right]$  the alternative $1/\left[s - M^2 + i \sqrt{s}\, \Gamma\right]$. We find essentially the same solution to the FESR, at five-loop order, for $s_0$ and the $d=4,6$ vacuum condensates provided the leptonic coupling $f_\rho$ is somewhat smaller, i.e. $f_\rho \simeq 3$. This will play no role at finite $T$, as we shall normalize all results to their $T=0$ values, i.e. we are essentially interested in finding the thermal behavior of ratios. These ratios are hardly distinguishable from those using Eq.(6).\\
\begin{figure}
\centering
\def\svgwidth{0.8\columnwidth}
\includegraphics[height=3.0in, width=3.5in]{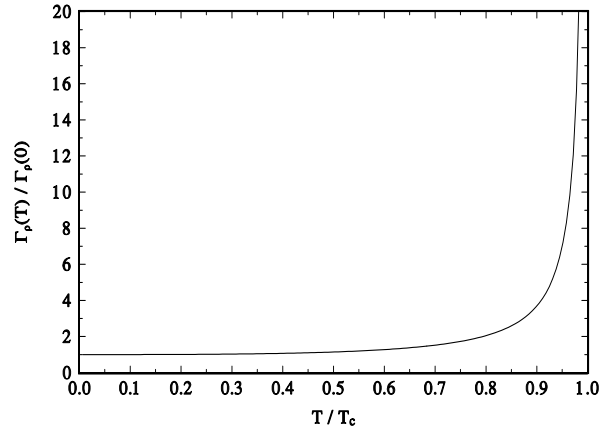}
\caption{{\protect\small{The normalized thermal behaviour of the $\rho$-meson width, for $T_c = 197 \;{\mbox{MeV}}$.}}}
\end{figure}
\begin{figure}
\centering
\def\svgwidth{0.8\columnwidth}
\includegraphics[height=3.0in, width=3.5in]{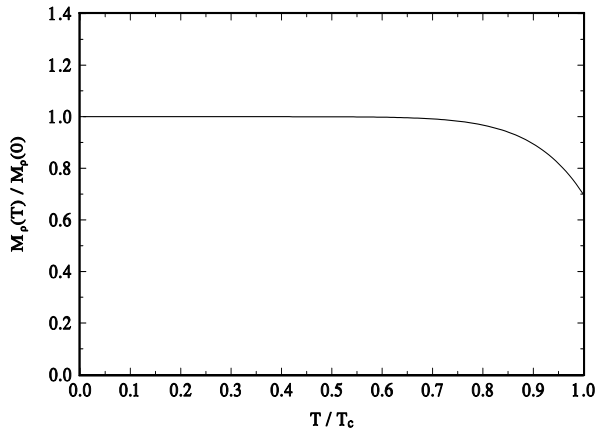}
\caption{{\protect\small{The normalized thermal behaviour of the $\rho$-meson mass, for $T_c = 197 \;{\mbox{MeV}}$.}}}
\end{figure}
\begin{figure}
\centering
\def\svgwidth{0.8\columnwidth}
\includegraphics[height=3.0in, width=3.5in]{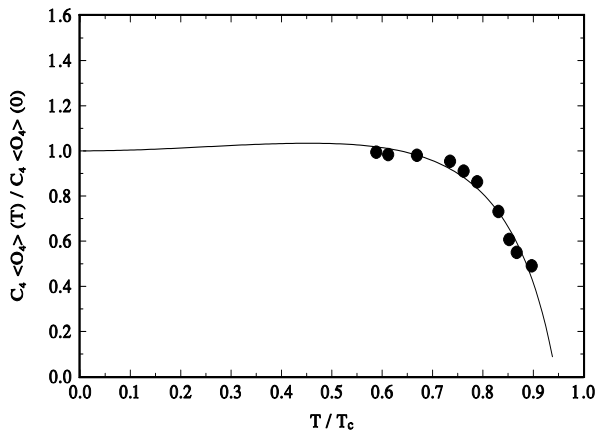}
\caption{{\protect\small{The normalized thermal behaviour of the dimension $d=4$ gluon condensate (solid curve), together with lattice QCD results \cite{LQCD} (solid circles) for $T_c = 197 \;{\mbox{MeV}}$.}}}
\end{figure}
\begin{figure}
\centering
\def\svgwidth{0.8\columnwidth}
\includegraphics[height=3.0in, width=3.5in]{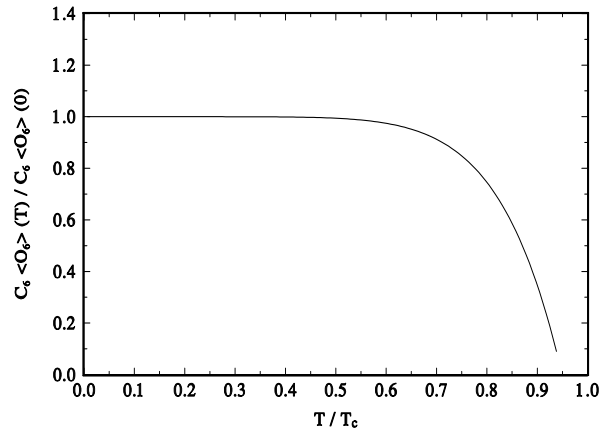}
\caption{{\protect\small{The normalized thermal behaviour of the dimension $d=6$ four-quark condensate, for $T_c = 197 \;{\mbox{MeV}}$.}}}
\end{figure}
The extension of the FESR program to finite temperature was first outlined in \cite{BS}. Field theory arguments in support of the validity of this extension were later given in \cite{OPET}. At finite $T$ there is an additional longitudinal structure in Eq.(1), but we shall consider the FESR for the transverse part. In the QCD sector one needs to restrict PQCD to the leading, one-loop level, as  the appearance of two scales in $\alpha_s(q^2,T)$, i.e. $\Lambda_{QCD}$ and $T_c$, remains an open problem (QCD sum rules approach $T_c$ starting from $T=0$, where PQCD is not valid). At this order there are two thermal contributions to the vector correlator, one in the time-like region ($q^2 > 0$) and one in the space-like region ($q^2 < 0$). In the static limit (${\bf{q}} \rightarrow 0$) these terms are
\begin{equation}
{\mbox{Im}} \, \Pi^{+}(\omega,T) = \frac{1}{4\,\pi} \left[1 - 2 \, n_F\left(\frac{\omega}{2 T}\right)\right]\;,
\end{equation}
for the time-like contribution, and
\begin{eqnarray}
{\mbox{Im}} \, \Pi^{-}(\omega,T) &=& \frac{4}{\pi} \; \delta(\omega^2) \; \int_0^\infty \, y \; n_F\left( \frac{y}{T}\right) \; dy \nonumber \\ [.3cm]
&=&\frac{\pi}{3}\; T^2 \, \delta(\omega^2)\;, \label{ST}
\end{eqnarray}
in the space-like region, where $n_F(z) = 1/(1 + e^{z})$ is the Fermi thermal function, and the chiral limit was assumed. The vacuum condensates develop a $T$-dependence which can be obtained from the sum rules themselves, or by resorting to lattice QCD (LQCD) determinations. A non-gauge invariant, non-zero dimension $d=2$ term in the OPE only appears at high temperatures \cite{C2T}, beyond the domain being normally explored with QCDSR so that it can be safely neglected.\\
In the hadronic sector the leptonic coupling, the mass, and the width of the $\rho$-meson entering Eq.\eqref{BW} become temperature dependent. In addition, there is a hadronic contribution in the space-like region due to the coupling of the vector current to two pions in the thermal bath, and given by
\begin{equation}
	\frac{1}{\pi}{\mbox{Im}}\, \Pi^{-}|_{HAD}(\omega,T) = \frac{2}{3\pi^2}\delta(\omega^2)\int_0^\infty y\, n_B\left(\frac{y}{T}\right)dy \;,
\end{equation}
where $n_B(z) = 1/(e^{z}-1)$ is the Bose thermal function.
\section{RESULTS AND CONCLUSIONS}
Given that there are only three leading FESR, and a few more parameters, the following strategy has been adopted. First, results for the  quark condensate from LQCD \cite{LQCDqq} was used as a first approximation to $C_6 \langle {\cal{O}}_6\rangle (T)$ in vacuum saturation. Next, we took guidance from the temperature dependence of leptonic couplings, masses, and widths determined in other light- and heavy-light-quark hadronic channels. We then explored the six-dimensional parameter space using this information, seeking a solution to the three FESR that would make physical sense, i.e. that would agree with expectations as developed in other channels. This means a decreasing $s_0(T)$, $f_\rho(T)$, $C_4 \langle {\cal{O}}_4\rangle (T)$, and $C_6 \langle {\cal{O}}_6\rangle (T)$, together with an increasing $\Gamma(T)$ with increasing $T$. A scanning of the parameter space shows a welcome highly peaked structure with almost overlapping solutions involving two {\it critical} temperatures. One for deconfinement, $T_c$, pertaining to the vanishing of $s_0(T)$, $f_\rho(T)$, and $C_4 \langle {\cal{O}}_4\rangle (T)$, and to the divergence of $\Gamma_\rho(T)$. And another temperature for the vanishing of the four-quark condensate, $T^*_q$, which turns out to be essentially the same as that for the vanishing of the gluon condensate, and some 5\% lower than $T_c$. This set of solutions require the $\rho$-mass to decrease with increasing $T$, although it remains non-zero at $T=T_c$. However, since the resonance width diverges at this temperature, the $\rho$-meson is no longer seen in the spectrum. The FESR cease to have solutions close to the deconfinement temperature, $T/T_c \simeq 0.90-0.95$, as found in previous analyses of this channel \cite{BS}-\cite{CAD1}, as well as in the axial-vector channel \cite{CAD6}-\cite{Axial}. The results are shown in Figs. 1-6, and correspond to the following analytical expressions
\begin{equation}
	\Gamma_\rho(T) = \frac{\Gamma_\rho(0)}{1- (T/T_c)^a}\;, \label{GAMMAf}
\end{equation}
where $a = 3$, and $T_c = 197 \,{\mbox{MeV}}$,
\begin{equation}
C_6\langle{\hat{O}_6}\rangle(T) = C_6\langle{\hat{O}_6}\rangle(0)\left[1- (T/T_q^*)^b\right]\;, \label{C6f}
\end{equation}
with $b=8$, and  $T_q^* = 187 \;{\mbox{MeV}}$, and
\begin{equation}
	M_\rho(T) = M_\rho(0)\left[1- (T/T_M^*)^c\right]\;, \label{MASS}
\end{equation}
where $c=10$, and $T_M^* = 222 \;{\mbox{MeV}}$, constrained to satisfy $T_M^* > T_c$. The slight $ 5 \%$ difference between $T_c$ and $T_q^*$ is well within the accuracy of the method. A change in the values  of the parameters $a, b, c$ in Eqs.\eqref{GAMMAf}-\eqref{MASS} affects the behaviour of $s_0(T)$, $f_\rho(T)$, and $C_4\langle{\hat{O}_4}\rangle(T)$. In order to retain the qualitative behaviour of the full six quantities, the parameters $a, b, c$ are restricted to changes not greater than $\pm \,30\,\%$, $\pm \,50\,\%$, and $\pm \,30\,\%$, respectively. The temperature  $T^*_q$ is rather tight, with a maximum allowed change of $\pm \, 3 \; {\mbox{MeV}}$, while $T^*_M$
could vary in the range $T^*_M = 210 - 240 \;{\mbox{MeV}}$.\\
 A fit to the results for the remaining three parameters gives $s_0(T)/s_0(0) = 1 - 0.5667 \;(T/T_c)^{11.38} - 4.347\; (T/T_c)^{68.41}$, $C_4\langle{\hat{O}_4}\rangle(T)/C_4\langle{\hat{O}_4}\rangle(0) = 1 - 1.65 \;(T/T_c)^{8.735} + 0.04967\; (T/T_c)^{0.7211}$, and $f_\rho(T)/f_\rho(0) = 1 - 0.3901 \;(T/T_c)^{10.75} + 0.04155 \;(T/T_c)^{1.269}$, corresponding to $T_c = 197\; {\mbox{MeV}}$.\\
The behavior of $s_0(T)$, Fig. 1, is  somewhat similar to the recent  result in the axial-vector channel \cite{CAD6}. While the hadronic spectral function is very different in these two channels, PQCD is chiral-symmetric (in the chiral limit). This result is pointing to an approximate universality of the deconfinement transition in light-quark systems. The  different behaviour close to the critical temperature can be traced to the contribution of the quark condensate (equivalently the pion decay constant) in the axial-vector channel, which is absent in the vector correlator (at $d=4$ the term $m_q \langle \bar{q} q \rangle$ is negligible).
The thermal width of the $\rho$-meson, Fig. 3, exhibits a dramatic increase of roughly a factor 20 near $T_c$. Its functional form, Eq.\eqref{GAMMAf}, should be of use in current experiments measuring dimuon production in heavy ion collisions \cite{BROADEXP}.
The result for the thermal gluon condensate, Fig. 5, is in good agreement with LQCD \cite{LQCD}, and the four-quark condensate, Fig. 6, is compatible with the behaviour of $|\langle \bar{q} q\rangle(T)|^2$, albeit with a coefficient different from that in the vacuum saturation approximation. This coefficient, though, cancels out in the ratio.
\section{Acknowledgments} 
This work has been supported in part by NRF (South Africa), FONDECYT (Chile) 1095217, 1120770, Proyecto Anillos (Chile) ACT 119 , DGAPA-UNAM (Mexico) under grant No. IN103811, and CONACyT (Mexico) under grant No. 128534. 


\begin{thebibliography}{99}
\bibitem{BS} A.I. Bochkarev and M.E. Shaposnikov, Nucl. Phys. B {\bf {268}}, 220 (1986).
\bibitem{Rho} R.J. Furnstahl, T. Hatsuda and S.H. Lee,  Phys. Rev. D {\bf 42}, 1744 (1990);
C. Adami, T. Hatsuda and I. Zahed, Phys. Rev. D {\bf 43}, 921 (1991); C. Adami and I. Zahed, Phys. Rev. D {\bf 45}, 4312 (1992); T. Hatsuda, Y. Koike and S.-H. Lee, Phys. Rev. D {\bf 47}, 1225 (1993); {\it ibid.} Nucl. Phys. B {\bf 394}, 221 (1993); Y. Koike, Phys. Rev. D {\bf 48}, 2313 (1993). 
\bibitem{CAD1} C. A. Dominguez and M. Loewe, Z. Phys. C {\bf 51}, 69 (1991).
\bibitem{CAD2} C. A. Dominguez, Deconfinement and chiral-symmetry restoration in finite temperature QCD, University of Cape Town Report No. UCT-TP-218/94 (unpublished); for an abridged version see: American Institute of Physics Conference Proceedings {\bf 342}, 383 (1995).
\bibitem{REVIEW} For a recent review see e.g. P. Colangelo and A. Khodjamirian, in {\it At the Frontier of Particle Physics: Handbook of QCD}, M. Shifman, ed. (World Scientific, Singapore 2001), Vol. 3, pp. 1495-1576.
\bibitem{Pich} For a review see S. Bethke {\it {et al.}}, Workshop on Precision Measurements of $\alpha_s$, arXiv: 1110.0016. 
\bibitem{CAD3} For a review see C. A. Dominguez,  Mod. Phys. Lett. A {\bf 26}, 691 (2011).
\bibitem{CAD4a} C. A. Dominguez and  M. Loewe, Z. Phys. C {\bf 58}, 273 (1993); C. A. Dominguez, M. S. Fetea and M. Loewe, Phys. Lett.  B {\bf 387}, 151 (1996).
\bibitem{CAD4}C. A. Dominguez and  M. Loewe, Phys. Lett. B {\bf 481}, 295 (2000); C. A. Dominguez, M. Loewe and J. S. Rozowsky, Phys. Lett. B {\bf 335}, 506 (1994); C. A. Dominguez, M. S. Fetea and M. Loewe, Phys. Lett.   B {\bf 406}, 149 (1997); C. A. Dominguez, M. Loewe and  C. van Gend, Phys. Lett.  B {\bf 429}, 64 (1998); {\it ibid} B {\bf 460}, 442 (1999);  
C. A. Dominguez, M. Loewe, and J. C. Rojas, J. High Energy Phys. {\bf {08}}, 040 (2007); E. V. Veliv and G. Kaya, Eur. Phys. J. C {\bf{63}}, 87 (2009).
\bibitem{BROAD}C. A. Dominguez and M. Loewe, Phys. Lett. B {\bf {233}}, 201 (1989). 
\bibitem{DIMUON}C. A. Dominguez and M. Loewe,  Z. Phys. C  {\bf {49}}, 423 (1991); For a recent analysis see R. Rapp, AIP Conf. Proc. 1322, 55 (2010).
\bibitem{BROADEXP} A. L. S. Angelis {\it et al.}, HELIOS-3 Collaboration, Eur. Phys. J. C {\bf 13}, 433 (2000); M. C. Abreu {\it et al.}, NA38/NA50 Collaboration, Nucl. Phys. A {\bf 698}, 539 (2002);
D. Adamova {\it et al.}, CERES Collaboration, Phys. Lett. B {\bf 666}, 425 (2008); G. Agakichiev {\it et al.}, CERES Collaboration, Eur. Phys. J. C {\bf 41}, 475 (2005); R. Arnaldi {\it et al.}, NA60 Collaboration, Eur. Phys. J. C {\bf 61}, 711 (2009).
\bibitem{CAD5} C. A. Dominguez, M. Loewe, J. C. Rojas, and Y. Zhang, Phys. Rev. D {\bf {81}}, 014007 (2010); {\it{ibid.}} D {\bf {83}}, 034033 (2011).
\bibitem{LQCD} M. Cheng {\it et al.}, Phys. Rev. D {\bf 77}, 014511 (2008).
\bibitem{sigma}C. A. Dominguez, M. Loewe, and J. C. Rojas, Phys. Lett. B {\bf 320}, 377 (1994).
\bibitem{CHPTT} J. Goity and H. Leutwyler, Phys. Lett. B {\bf 228}, 517 (1989); H. Leutwyler and A. V. Smilga, Nucl. Phys. B {\bf 342}, 302 (1990).
\bibitem{CAD6} C. A. Dominguez, M. Loewe, and Y. Zhang, Phys. Rev. D {\bf 86}, 034030 (2012).
\bibitem{Axial} A. Barducci {\it{et al.}}, Phys. Lett. B {\bf {244}}, 311 (1990); A. Ayala {\it{et al.}}, Phys. Rev. D {\bf {84}}, 056004 (2011).
\bibitem{taucond} C. A. Dominguez and K. Schilcher, J. High Energy Phys. {\bf {0701}}, 093 (2007).
\bibitem{eecond} S. Bodenstein, C. A. Dominguez, S. Eidelman, K. Schilcher, and H. Spiesberger, J. High Energy Phys. {\bf 01}, 039 (2012).
\bibitem{VS1} A. Gomez Nicola, J. R. Pelaez, and J. Ruiz de Elvira, Phys. Rev. D {\bf 82}, 074012 (2010).
\bibitem{VS2}A. Gomez Nicola, J. R. Pelaez, and J. Ruiz de Elvira, arXiv: 1210.7977.
\bibitem{PDG} J. Beringer {\it{et al.}} (Particle Data Group) Phys. Rev. D \textbf{86}, 010001 (2012).
\bibitem{OPET} C. A. Dominguez and M. Loewe, Phys. Rev. D {\bf {52}}, 3143 (1995).
\bibitem{C2T}E. Megias, E. Ruiz-Arriola, and L. L. Salcedo, Phys. Rev. D {\bf{81}}, 096009 (2010).
\bibitem{LQCDqq}A. Bazavov {\it et al.}, \prd {\bf 80}, 014504 (2009); M. Cheng {\it et al.},
\prd {\bf 81}, 054504 (2010).
\end{thebibliography}
\end{document}